\documentclass{aa}\usepackage{graphicx}
\def\OMEGA0{\Omega_{\rm m,0}}
\def\lambda0{\Omega_{\Lambda,0}}

\def\kms{\,\rm km\,{s}^{-1}}

\def\mpc{\,{\rm Mpc}}
\def\LCDM{\Lambda{\rm CDM}}

\def\rms{{rms}}
\def\Rcut{R_{\rm cut}}
\def\beq{\begin{equation}}
\def\eeq{\end{equation}}

\begin{document}

\title{Are Great Disks Defined by Satellite Galaxies in Milky-Way Type Halos Rare
in $\Lambda$CDM?}
\author{X. Kang\inst{1} \and S. Mao\inst{2}\and
L. Gao\inst{3}\and Y. P. Jing\inst{1}
}
\offprints{ X. Kang}
\mail{kangx@shao.ac.cn}

\institute{ Shanghai Astronomical Observatory; the Partner Group
of MPA, Nandan Road 80, Shanghai 200030, China;
\and
 University of Manchester, Jodrell Bank Observatory,
  Macclesfield, Cheshire SK11 9DL, UK;
\and
 Max-Planck Institute for Astrophysics,
  Karl-Schwarzschild-strasse 1, Garching 85748, Germany
}

\date{Received; accepted}

\abstract{ We study the spatial distribution of satellite galaxies
by assuming that they follow the dark matter distribution. This
assumption is supported by semi-analytical studies based on
high-resolution numerical simulations.  We find that for a
Milky-Way type halo, if only a dozen satellite galaxies are
observed, then they can lie on a ``great'' disk with an rms height
of about 40\,kpc. The normal to the plane is roughly isotropic on
the sky. These results are consistent with the observed properties
of the satellite galaxies in the Milky Way.  If, however, the
satellite galaxies follow the distribution of substructure
selected by present mass, then great disks similar to the one in the
Milky Way are rare and difficult to reproduce, in agreement with
the conclusion reached by Kroupa et al. (2004). \keywords{
cosmology -- Galaxy: evolution -- Galaxy: halo -- galaxies: dwarf
-- galaxies: structure }}

\maketitle

\section{Introduction}

The Cold Dark Matter (CDM) structure formation model
has been successful in explaining many observations,
in particular the large-scale structure of the
universe (e.g. Peacock et al. 2001; Tegmark et al. 2004)
and the cosmic microwave background (e.g., Spergel et
al. 2003). However, on small scales, it appears to have problems
in reproducing the rotation curves of dwarf galaxies
and the satellite properties in Milky-Way type halos
(e.g., see Silk 2004 for a review). But it is still a matter of
debate whether these conflicts are
real and whether they can be resolved within the CDM framework.

The CDM model generically predicts the existence of substructures
(subhalos). In this model, larger structures form by merging of
smaller structures, and dense cores in these small structures
often survive the tidal force and manifest as substructures. These
substructures are most clearly seen in high-resolution numerical
simulations (e.g., Klypin et al. 1999; Moore et al. 1999; Ghigna
et al. 2000). If all these substructures host luminous galaxies,
then the number of satellite galaxies in a Milky-Way type halo can
reach several hundred, clearly exceeding the number of known
satellite galaxies in the Milky Way (e.g., Kauffmann et al. 1993).
But if only a small fraction ($\sim 10\%$) of the substructures form
stars during their evolution, then the number of observed
satellites can be reconciled with observations (Kravtsov et al. 2004). Whether the
internal kinematics of satellite galaxies are consistent with the
observations is still unclear (Stoehr et al. 2002; Kazantzidis et al. 2004).

Recently, Kroupa et al. (2004) highlighted the fact that the observed
spatial distribution of satellite galaxies in the Milky Way is distributed
in what they term as a ``great'' disk; its plane is almost perpendicular
to the plane of the Galactic disk. These properties are difficult to
understand if the satellite galaxies follow
the quasi-spherical distribution expected for the
substructures. Recently, Gao et al. (2004a; see also Springel et
al. 2001; Diemand, Moore \& Stadel 2004; Nagi \& Kravtsov 2005) combined semi-analytical
techniques and high-resolution numerical simulations to examine
the relation between substructures and (satellite) galaxies in
clusters of galaxies. They showed that satellite galaxies follow
roughly the same distribution as the underlying dark matter while the
substructures follow a much shallower density distribution in the
central part. In this paper, we
show that if the satellite galaxies follow
the same spatial distribution as dark matter, then
many of their puzzling spatial properties highlighted by
Kroupa et al. (2004) are easier to understand.

The outline of the paper is as follows. In \S2 we first describe the simulation
data we use, and then compare the spatial distribution of
satellite galaxies with observations, concentrating on the statistics used
by Kroupa et al. (2004). We discuss our results in \S3.
Throughout this paper, we adopt the ``concordance'' $\LCDM$
cosmology (e.g., Ostriker \&
Steinhardt 1995; Spergel et al. 2003 and references therein),
with a matter density parameter $\OMEGA0=0.3$, a cosmologically constant
$\lambda0=0.7$, a baryon density parameter $\Omega_{\rm b}=0.024h^{-2}$,
and we take the power-spectrum normalization $\sigma_8=0.9$. We write the
Hubble constant as $H_0=100\,h\kms\mpc^{-1}$ and adopt $h=0.7$.

\section{Method and Results}

The data we use in this paper is the high resolution halo
simulation of Jing \& Suto (2000). Four halos
on galactic-mass scales (around $4\times 10^{12}h^{-1}M_{\odot}$)
were selected from a cosmological simulation
with a  box size of $100h^{-1}\mpc$. They were then re-simulated using the
nested-grid PPPM code which was designed to simulate high-resolution
halos.  The force resolution is typically $0.4\%$ of the virial
radius, and each particle has mass about $6 \times 10^{6}h^{-1}M_{\odot}$
for the four galactic halos. We use the {\tt SUBFIND} routine of
Springel et al. (2001) to identify
the disjoint self-bound subhalos within these halos and all subhalos with more
than 10 particles are included in our analysis. Furthermore, in order
to check the distribution of satellite galaxies in galactic halos,
we use the galaxy catalog in
a cosmological simulation of box $25h^{-1}\mpc$ constructed by
Kang et al. (2004) using their semi-analytical model of galaxy formation.
The sub-galactic satellites in their simulation were resolved with
more than 10 particles, and each particle has mass of $7.7\times 10^{7}h^{-1}M_{\odot}$.

\begin{figure}
\resizebox{\hsize}{!}{\includegraphics{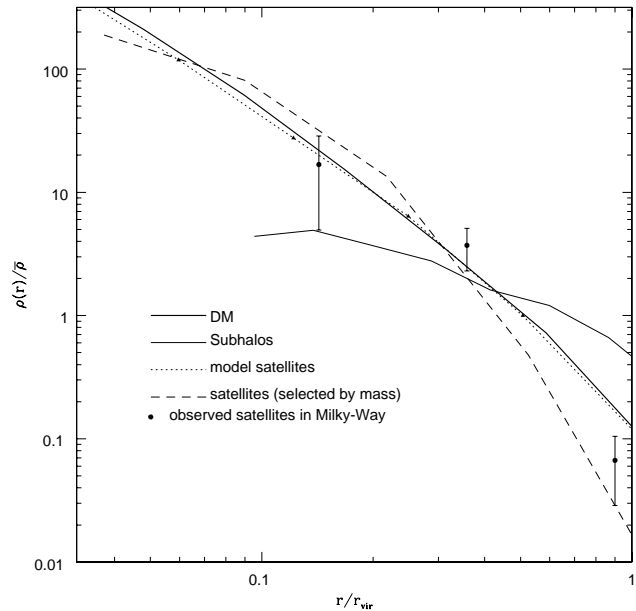}} \caption{ The
radial distribution of satellite galaxies. The solid circles are
the distribution of the 11 inner satellites observed in the Milky
Way (Mateo 1998; Kroupa et al. 2004); the Poisson error is
indicated for each data point. The thick solid line is the dark
matter profile while the thin solid line is for the substructures.
The dotted line with triangles is for the satellite galaxies
(bright than $M_{V}=-12$) in galactic halos predicted by the
semi-analytical model of Kang et al. (2004), and the dashed line
is for the sample selected by mass at accretion (see text). Here
the radius is normalized to the virial radius and the virial
radius of the Milky Way is taken to be 250\,kpc. } \label{Profile}
\end{figure}

It is known that in galaxy clusters the observed galaxy
distribution follows the underlying dark matter (Gao et al.
2004a), but it is less clear whether the observed satellite
galaxies in galactic mass halos follow the same distribution as
the dark matter. In Fig. \ref{Profile} we show the distribution of 
the observed satellites of the Milky Way and the comparisons with 
our simulation. The
dotted line with triangles are the model satellite galaxies (with
$M_{V}<-12$) in galactic halos in our simulation. The thick solid
line is the distribution of dark matter in the high-resolution
galactic halos. We found that the modeled satellite galaxies
follow the underling dark matter distribution very well. Also it
can be seen that the observed satellites roughly follow the dark
matter particles. The Kolmogorov-Smirnov (KS) test shows that
their distributions are consistent at a 60\% confidence
level. In contrast, the substructures with more than 10 particles
clearly differ from the dark matter or the observed satellites
distribution. In particular, they are not significantly centrally
concentrated and have a higher density than observation in the
outer part, as also found by others (De Lucia et al. 2004; Gao et al.
2004a; Diemand et al. 2004). Note that our subhalo population is
incomplete as we can only identify those with more than ten
particles ($>6\times 10^7h^{-1}M_\odot$). Increasing the mass resolution will
result in more subhalos, but the relative number density profile will
not change with the resolution as higher resolution primarily allows
the measurement of the density profile near the centre (Diemand et al.
2004).

If all the subhalos form stars then the predicted spatial distribution
of the satellites will be in conflict with the observed distribution.
Furthermore, many models of galaxy formation will
over-predict the number of dwarf satellite galaxies by an order of
magnitude (Kauffmann et al. 1993; Bullock et al. 2000; Somerville
2002; Benson et al. 2002, Kang et al. 2004). However, if there is a 
bias between the satellite galaxies and the subhalo population, then
this over-abundance problem can be alleviated (Taylor et al. 2003; Kravtsov et
al. 2004). For example, using high-resolution N-body simulations,
Kravtsov et al. (2004) showed that
if only a fraction ($\sim 10\%$) of the subhalos
with mass $\leq 10^{8} - 10^{9} M_{\odot}$ host luminous galaxies,
then the observed number of galactic satellites and their spatial
distribution can be reproduced. They found that these luminous
satellites are descendants of accreted massive halos ($\geq
10^{9}M_{\odot}$) at $z \geq 2$. These massive systems with
temperature larger than $10^{4}$K can have efficient gas cooling
and star formation during their evolutions. Motivated by this study,
we check whether satellites (in a Milky-Way type halo) 
selected in this way have a similar distribution as the observed satellites.
There are about 100 massive halos ($\geq
10^{9}h^{-1}M_{\odot}$) selected at accretion with redshift $z
\geq 2$. The resolution of our re-simulated halos is not high
enough for us to fully resolve the dwarf galaxies with mass around
$10^{7}M_{\odot}$. So we, instead, use the dynamical friction timescale given
by Navarro et al. (1995, see their eq. 2) to determine which
of the 100 most massive halos can survive in the final
galactic halo, and we find that 30 of them survived. In order to obtain 
their final positions, we identify the most-bound particles of these 
halos at accretion and the final positions of the survied halos at $z=0$ 
are tagged by these most-bound particles. This is reasonable as high-resolution 
N-body simulations have shown that the inner part of the halo will keep intact 
during the evolution (e.g. Springel et al. 2001). The radial distribution of 
these survived halos is plotted in Figure 1 with 
the dashed line. Clearly
the satellite galaxies selected by mass at accretion
have a similar distribution with the observed satellites in the
Milky Way.  The KS test shows that it also agrees with the
radial distribution of the dark matter particles at a
20\% confidence level. We also check if the satellites selected in this way 
have a similar shape with the dark matter particles. The number of satellites (30)  
is not large enough to define their shape accurately. Nevertheless,
we fit their distribution with 
a triaxial distribution and characterize their shape as the ratio between the minor and 
major axes. We then use Monte Carlo simulation to produce a distribution
of such ratios by selecting the satellites from the underlying 
dark matter particles. We found that the ratio of the satellites selected by mass 
at accretion lies at the $75\%$ percentile of the distribution. 
This agreement provides the justification to
select the satellites galaxies from the dark matter particles in
the re-simulated galactic halos. We also consider the case when
the satellite galaxies follow the distribution of the subhalos
($\geq 6 \times 10^{7}h^{-1}M_{\odot}$), but we will show
that their distribution is not consistent with the observations.

To examine the spatial distribution of satellites in
our simulated halos, we use Monte Carlo simulations to select the
satellites assuming they follow either the dark matter
distribution or the substructure distribution. For a full
comparison with the results of Kroupa et al. (2004), we select 11
random satellites in a sphere with radius of about 250 kpc from
the centre. In order to see how the results change with the number
of satellites, we also show the results by increasing the number
of satellites by a factor of 2, i.e., 22 satellites within the
same sphere.

\begin{figure}
\resizebox{\hsize}{!}{\includegraphics{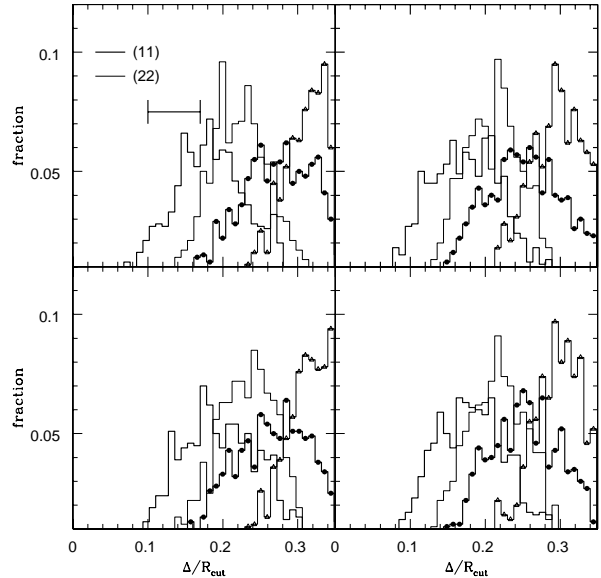}}
\caption{
The distribution of the rms of the scale height from a plane defined by satellite
galaxies obtained through many Monte Carlo realizations.  The four
panels are for four different galactic-sized halos.
The histograms without symbols show that expected if the satellite galaxies follow the
dark matter distribution while the histograms with symbols show those if satellites
follow the subhalo distribution. The thick solid histograms are for 11 satellite galaxies,
while the thin solid histograms are for 22 satellite galaxies. The thick
horizontal bar in the top left panel
shows the range of $\Delta/\Rcut$ for the observed satellites
when different numbers of satellite galaxies are used (see Table 1 in Kroupa et al. 2004).}
\label{Dr}
\end{figure}

Following Kroupa et al. (2004), we fit the selected satellites
with a plane by minimizing the \rms\, of the height. The thickness
of the plane is the minimum of the \rms\, of the height $\Delta$.
The ratio of $\Delta$ to the maximum distance ($\Rcut$) to the
satellites is used to characterize the thickness of the plane. In
Fig. \ref{Dr} we show the distribution of the characteristic
thickness $\Delta/\Rcut$ of the fitted plane from a large number
of Monte Carlo realizations. The four panels are for the four
different halos. In these plots the histograms without symbols are
the distributions of the satellites selected from the dark matter
particles and those with symbols are for the substructures. We
find that for dark matter samples, the thickness of the planes are
smaller than that in the samples of substructures. The average
thickness of the planes peaks at $\Delta/\Rcut \sim 0.17$. It can
also be seen that increasing the number of satellites also
increases the thickness of the planes. Kroupa et al. (2004) showed
that for the inner 11 satellites the characteristic thickness of
the plane is $\Delta/\Rcut \approx 0.1$, but this $\Delta/\Rcut$
value changes when one adopts a different number of satellites.
For example, for the 9 innermost satellites, the characteristic
thickness of the plane is 0.17. The thick horizontal bar in the top left
panel of Fig. \ref{Dr} indicates the range of $\Delta/\Rcut$ if one adopts 5-16
satellite galaxies in the Milky Way.
Clearly it is difficult to produce a value as
small as the observed one if the satellite galaxies follow the
substructure distribution. In contrast, if the satellite galaxies
follow the dark matter distribution, it is not difficult to have
the observed low value of $\Delta/\Rcut$. It is easy to show that
for a given spherical density profile $\rho \propto r^{-n}$, the
thickness of the plane as defined here is given by
\begin{equation}
{\Delta \over \Rcut} = \sqrt {\frac {3-n} {3(5-n)}},~ n<3.
\end{equation}
The above equation is valid when all the particles in the
sphere are selected. Notice also that when the power-law index $n \rightarrow
3$, the thickness
goes to zero. This is because the mass and hence the particle number
diverge at the origin when $n\rightarrow 3$. The plane determined  by
these (infinite) number of particles around the origin will
have formally zero thickness. 

>From Fig. \ref{Profile} we see that the
substructures have a shallower slope (i.e., a smaller value of
$n$) at the centre, so most of them are in the outer part of the halo,
which makes the fitted plane thicker. The observed satellites, in
contrast, follow a steeper slope than the substructures, so
satellites are more likely in the inner part of the halo and hence
make the plane more prominent. In fact the thickness also
depends on the number of particles selected. Clearly a plane
with only 3 particles will have zero thickness while a plane with infinite
number of particles will have a thickness given by eq. (1). 
This is why a larger characteristic thickness is found when we increase
the satellite number from 11 to 22 in Fig. \ref{Dr}.

\begin{figure}
\resizebox{\hsize}{!}{\includegraphics{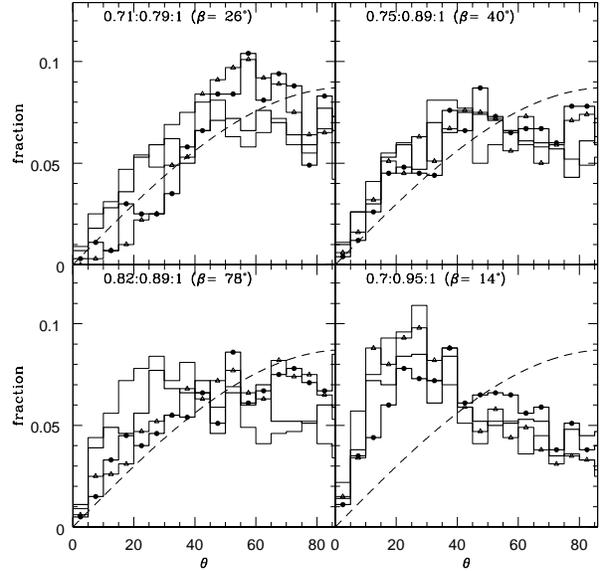}}
\caption{
The distribution of the rms of the angle (in degrees) between the plane defined by satellite
galaxies and the plane defined by the major and medium axes in the
triaxial halo model (see text). The dashed line is the prediction if
the angle between the satellite plane
and the minor axis of the
halo is randomly distributed on the sky. The other line symbols are the same as
in Fig. \ref{Dr}. In each panel we label the ratio of the minor,
medium, maximum axes. In the parentheses we indicate the angle between the angular momentum vector
and the minor axis of the triaxial model.
}
\label{angle}
\end{figure}

Kroupa et al. (2004) points out that the fitted plane of the Milky
Way satellites is almost perpendicular to the Galactic disk; the normals
to these two planes have an angle of approximately $75^\circ$.
Unfortunately it is not clear how we should define the disk plane
in our simulated galaxies. There are at least two ways of defining
the disk plane. One is simply to identify the total angular
momentum vector as the normal to the disk plane. The second way is
that we can fit the density profile of the halos with a triaxial
distribution, obtain the three principal vectors (${\bf a}$, ${\bf
b}$ and ${\bf c}$, $a \leq b \leq c$), and identify the disk plane
as that defined by the major and medium axes. These two
definitions would be the same if the minor axis is parallel to the
angular momentum vector. While there is a statistical correlation
between these two, the
scatter is quite large (e.g., Faltenbacher et al. 2002). In Fig.
\ref{angle} we show the angle ($\beta$) between the total angular
momentum vector and the minor axis. Note that
we use all the particles enclosed within the virial radius to obtain the total
angular momentum vector and the triaxial density profile.
As one can see, the alignment for three halos is within
$40^\circ$, but for the halo in the lower left panel,
$\beta=78^\circ$, i.e., the minor axis is almost perpendicular to
the total angular momentum vector. The identification of the disk
plane is further complicated by the fact that it is not clear
how the angular momentum of the (baryonic) disk is
related to the total angular momentum of the dark matter in
N-body/hydro-dynamical simulations. Chen et al. (2003) showed that
there is a large scatter between these two.
\begin{figure}
\resizebox{\hsize}{!}{\includegraphics{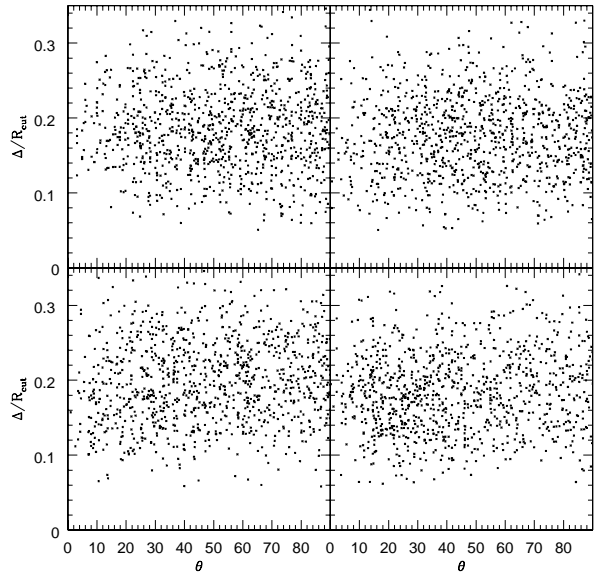}}
\caption{
A scatter plot of $\theta$ vs $\Delta/\Rcut$ for the satellites selected 
from the dark matter particles. $\theta$ is the same as that in Fig.\ref{angle}.
}
\label{scatter}
\end{figure}

In Fig. \ref{angle} we show the distribution of the angle ($\theta$)
between the normal vector to the fitted plane and the minor axis ${\bf
  a}$, assuming that the disk plane is the same as the plane defined
by the major and medium axes. 
In each panel the ratio of a:b:c is also labeled.
For two randomly oriented normal vectors, the probability distribution
for the angle between them is $\propto \sin\theta$, which is
shown as the dashed line in Fig. \ref{angle}. Compared with the
random distribution, the satellites are preferentially located
in a plane defined by the major and medium principal axes.
If we select only 11 satellite galaxies, then
the probabilities for the four halos to have an angle $\theta$ larger
than $60^\circ$ ($80^\circ$) are between 28-41\% (9\%-14\%), respectively.
So the probability for the fitted satellite plane to have a large angle
with the galaxy disk is not negligible. On the other hand, if
we identify the galaxy
disk as that given by the angular momentum vector, these corresponding
probabilities become even larger, $28\%-70\%$ (9\%-26\%) for
$\theta>60^\circ$ ($\theta>80^\circ$).

The above results show that there is some probability for the
selected satellite galaxies to lie in a plane and some probability
for the plane to be perpendicular to the disk. But it is important to
examine the joint distribution of $\Delta/\Rcut$ and the angle $\theta$.
Fig. \ref{scatter} shows the scatter plot between $\Delta/\Rcut$ and the angle $\theta$
for the satellites from dark matter distribution (the scatter plot
for substructures is similar, but of course with larger values of $\Delta/\Rcut$).
 As we can see, there is no correlation between them, for a given thickness of the
plane, it has roughly the same probability to have large and small
$\theta$ values. The probability for having $\theta>60^\circ
(80^\circ)$ and $\Delta/\Rcut<0.17$ ranges from 10\% to 17\% (from 3\% to
6\%) for the four different haloes. 
This means that it is not rare for the observed plane to have large inclinations
with the disk.

Kroupa et al. (2004) have analyzed the cumulative distribution of
$\cos \omega$ for the Milky Way satellites, where $\omega$ 
is the angle between the normal to the great disk and the vector
connecting the point on the great disk that is closest to the 
Galactic centre to the position of the satellite. As is designed, the
distribution of $\cos \omega$ is closely related to the thickness of
the great disk. They compared this distribution with that of $10^5$
points derived from a spherical power-law density distribution
$r^{-n}$. Using the KS test, they found the likelihood, $P_{\rm ks}$,
that the observed Milky Way distribution is derived from the $r^{-n}$ model
is only 0.005 for $0<n<2.3$. Taken at face value, this seriously challenges
the assumption that the Milky Way satellites are consistent with an isotropic
parent distribution.

We have repeated their analysis for the Milky Way, and found that $P_{\rm ks}$ is
$\sim 0.01 $ if the distribution of $|\cos\omega|$ is used, 
consistent with Kroupa et al. (2004), who used the same quantity
although they paper appears to imply a different quantity,
$\cos\omega$, was used (Kroupa 2005, private communication).
But there is no obvious reason why one should use
$|\cos\omega|$ instead of $\cos\omega$. The distribution of
$\cos\omega$ should equally well describe the thickness of the great
disk. When we use $\cos\omega$, we found that $P_{\rm ks}$ increases
dramatically to $\sim 20\%$ for the Milky Way satellites, which 
implies that the great disk is compatible with the hypothesis that the 
satellites are drawn from an isotropic parent population.

We randomly select 11 dark matter particles from the first galactic
halo, and repeat the above KS test for the simulation satellites. We
have made many realizations, and found that there is a 32\% (8.8\%)
probability that the $P_{\rm ks}$ of the simulated satellites is smaller
than that of the observed $\cos \omega$ ($|\cos\omega|$)
distribution for the Milky Way. About 40\% (19\%) of the simulated populations have
$P_{\rm ks}< 0.1$ for the $\cos \omega$ ( $|\cos\omega|$) distributions.
These results have two important implications. First,
the KS probability using $|\cos\omega|$ (or $\cos\omega$) should not be directly 
interpreted as the probability whether the satellites are consistent
with being drawn from a spherical $r^{-n}$ density distribution,
otherwise the probability may be under-estimated.
The reason is that the great disk is determined using the positions of all the
satellite galaxies, as a result the $\cos\omega$ values based on the great disk
are no longer independent of each other. This makes the KS probability
misleading as it assumes that the $\cos\omega$ values are independent
data points (e.g., Lupton 1993).
Here is an extreme example that illustrates this point clearly. If only 3
simulated satellites are selected from a spherical $r^{-n}$ distribution, then
$\cos\omega$ is zero for all the three satellites. One finds
that the KS probability defined above is very small ($\sim 0.13\%$). This low
probability is clearly misleading as this population
is drawn from an isotropic power-law density distribution. 
Second, our results 
clearly show that there is a 32\% (or 8.8\%) probability to reproduce
the observed $\cos\omega$ (or $|\cos\omega|$) distribution
in the $\Lambda$CDM halos, thus it is not so rare to expect
great disks similar to the one in the Milky Way in the $\Lambda$CDM model.

\section{Discussion}

The presence of dark matter substructures is a generic
prediction of the $\LCDM$ model. However, as pointed out by many
authors (e.g., Springel et al. 2001; Gao et al. 2004a,2004b; Nagi
\& Kravtsov 2005), the correspondence between dark matter substructures and
luminous satellite galaxies is not simple. This is because the
stellar mass of a galaxy may be primarily determined when it is
first formed. In contrast, the dark matter substructure may evolve
quite differently due to dynamical friction and other processes at
later times. In fact, in the study of Gao et al. (2004a), they
find $\sim 50\%$ of satellite galaxies have no corresponding dark matter
subhalos.

If the satellite galaxy distribution follows the dark matter
distribution, our Monte Carlo simulations demonstrate that the
spatial distribution of satellite galaxies can be better accommodated
within the $\LCDM$ cosmogony. In particular the fact that the
Milky Way satellite galaxies are distributed in a great disk with
its plane almost perpendicular to the stellar disk is not as rare
as one naively expects. In particular, for the halo shown in the
bottom left panel of Fig. \ref{angle}, as the angular momentum
vector of the halo is almost perpendicular to minor axis, it has
the largest probability, $>26\%$ (or 14\%) for the fitted plane to
have an angle larger than $80^\circ$ with the disk if we identify
the disk plane as that given by the total angular momentum vector
(or the minor axis in the triaxial model).

In contrast, if the satellite galaxies follow the distribution of
the substructures selected by the present-day mass 
then we find that the satellite galaxy properties will be
difficult to match because the substructures have a flatter 
slope with $n \sim 0$. But if only $\sim 10\%$ percent of the substructures 
host luminous galaxies, the observed distribution of the satellite galaxies 
can be recovered, as also shown by Kravtsov et al. (2004). It remains, however,
somewhat puzzling why the
observed ``great `` disk is almost perpendicular to the disk. As we have
shown at the end of \S 2, the probability of having a thin great disk perpendicular to
the Galactic plane is a few percent to $\sim 17\%$. In fact,
in our simulations, we find the
the fitted plane of satellite galaxies have a larger 
probability to lie in the disk defined by the major and medium axes in the
triaxial model. If the baryonic disk is in the same plane, then
the great disk in the Milky Way is most easily
explained if the Galactic plane has a large inclination
with the plane defined by the the major and medium axes of the dark
matter distribution. There is currently no observational evidence for this.

Unfortunately, the comparison between our simulations and observations
is not direct. Theoretically, with our collisionless, dark matter only
simulations, it is not clear how to relate 
the dark matter properties (such as the angular momentum vector) to
baryonic disk properties.
Furthermore, the dark matter halo and satellite galaxy profiles will be affected (perhaps
adiabatically, e.g., Mo, Mao \& White 1998) by the assembly of the baryonic disk. Furthermore, the
presence of the disk may force subhalos into roughly coplanar orbits
(Meza et al. 2004) and makes it more difficult to have the satellite
great disk perpendicular to the stellar disk. 
Observationally, it is not clear all the satellite galaxies in the Milky
Way have been found. In particular, some satellite galaxies in low Galactic latitudes may
have been missed (Willman et al. 2004). The addition of any such satellites
(if at large distances) will make the great disk in the Milky Way much thicker.
It appears that the issue of satellite galaxies can only be fully resolved 
when more observations and high-resolution hydrodynamical simulations become available.

\begin{acknowledgements}

We thank the referee, Dr. Juerg Diemand, for a helpful report and Dr. P. Kroupa
for clarifications.
SM and LG thank the hospitalities of Shanghai Astronomical Observatory
during several visits. SM acknowledges the financial support of Chinese Academy
of Sciences and the EU ANGLES research network. The research is supported
by NKBRSF (G19990754), by NSFC (Nos. 10125314, 10373012), and by Shanghai
Key Projects in Basic research (No. 04jc14079).  XK is supported in part
by NSFC (10203004).
\end{acknowledgements}

\end{document}